\documentclass[aps,prl,twocolumn,amsmath,amssymb,groupedaddress,nobibnotes,showpacs]{revtex4}

\usepackage{graphicx}
\usepackage{dcolumn}
\usepackage{bm}
\usepackage{amsmath}
\usepackage{amssymb,amsthm}
\usepackage{epsfig,color}
\usepackage[sans]{dsfont}

\definecolor{nblue}{rgb}{0.2,0.2,0.7}
\definecolor{ngreen}{rgb}{0.2,0.6,0.2}
\definecolor{nred}{rgb}{0.7,0.2,0.2}
\definecolor{nblack}{rgb}{0,0,0}

\newcommand{\tr}{\text{tr}}

\renewcommand{\H}{\mathcal{H}}

\def\x{\mathrm{ex}}

\def\a{\mathrm{ave}}
\def\w{\mathrm{min}}

\def\V{\mathcal{V}}

\def\E{\mathcal{E}}

\def\N{\mathcal{N}}

\def\tr{\mbox{tr}}
\def\bea{\begin{eqnarray}}
\def\eea{\end{eqnarray}}

\begin{document}

\title{Gate Fidelities, Quantum Broadcasting, and Assessing Experimental Realization}

\author{Hyang-Tag Lim}
\affiliation{Department of Physics, Pohang University of Science and Technology (POSTECH), Pohang, 790-784, Korea}

\author{Young-Sik Ra}
\affiliation{Department of Physics, Pohang University of Science and Technology (POSTECH), Pohang, 790-784, Korea}

\author{Yong-Su Kim}
\affiliation{Department of Physics, Pohang University of Science and Technology (POSTECH), Pohang, 790-784, Korea}

\author{Yoon-Ho Kim}
\affiliation{Department of Physics, Pohang University of Science and Technology (POSTECH), Pohang, 790-784, Korea}

\author{Joonwoo Bae}
\email{bae.joonwoo@gmail.com}
\affiliation{School of Computational Sciences, Korea Institute for Advanced Study, Seoul, 130-012, Korea}

\date{\today}


\begin{abstract}
We relate gate fidelities of experimentally realized quantum operations to the broadcasting property of their ideal operations, and show that the more parties a given quantum operation can broadcast to, the higher gate fidelities of its experimental realization are in general. This is shown by establishing the correspondence between two operational quantities, quantum state shareability and quantum broadcasting. This suggests that, to assess an experimental realization using gate fidelities, the worst case of realization such as noisy operations should be taken into account and then compared to obtained gate fidelities. In addition, based on the correspondence, we also translate results in quantum state shareability to their counterparts in quantum operations.


\end{abstract}



\maketitle



While quantum systems are processed to perform information tasks, building blocks to control and manipulate quantum states of interest are characterized by \emph{positive and completely positive} maps between quantum states. This immediately gives fundamental limitations on quantum state manipulation, and consequently \emph{approximate quantum operations} are naturally followed. Interestingly, approximate operations can be found in important quantum information applications: for instance, approximate quantum cloning implements optimal interaction to learn about given quantum systems \cite{ref:no-go}, or approximate operations to the partial transpose can be used to detect entanglement \cite{ref:he}. 

This introduces efficient implementation of approximate quantum operations as one of major challenges for quantum information applications, e.g. \cite{ref:lim}. As successful realization is closely related to achieving what is aimed by ideal quantum operations, it is thus important how an experimental realization of approximate operations is assessed. Gate fidelities that have been applied to unitary transformations \cite{ref:horodecki} \cite{ref:nielsen}, can be extended to general quantum operations since by gate fidelities it is quantified how faithful an experimental realization is. However, we point out that once gate fidelities are computed, interpretation to the obtained values is not clear. For instance, suppose that gate fidelities of experimentally realized operations with respect to the ideal one are obtained about $0.99$, which might claim very good performance. But, does it really mean that the ideal quantum operation has been faithfully implemented? As we will show later, what if a complete noise operation such as the depolarization can also achieve about $0.98$ in gate fidelities? In fact, gate fidelities can have different ranges according to given quantum operations.


In this work, motivated by the questions addressed in the above, we provide an analysis on and explain ranges that gate fidelities have. Namely, we establish the correspondence between the broadcasting property of quantum operations and gate fidelities of realized operations: the more parties an ideal operation can broadcast to, the higher values gate fidelities of its experimental realizations have in general, regardless of the quality of actual realizations. This is shown by proving that from the isomorphism between states and operations in Ref. \cite{ref:C-J}, \emph{the broadcasting property of quantum operations corresponds to the shareability of quantum states}. We then translate the hierarchy among shareable quantum states into quantum operations, and derive a general lower bound to gate fidelities. This provides both quantitative and qualitative interpretations to gate fidelities: i) by comparing obtained gate fidelities to its lowest bound, an experimental realization can quantified, and ii) using the broadcasting property, obtained gate fidelities can explain how noisier the realization is compared to the ideal one. Through the correspondence, we also translate results in quantum state shareability into properties of quantum broadcasting operations.



Let us begin with identifying gate fidelities that we are going to consider. Throughout the paper, we consider a quantum operation $\E$ which approximates an operation $\V$ that might not be possible in quantum theory. For instance, $\E$ is an optimal quantum cloning that approximates perfect cloning operation $\V$. Let $\E_{\x}$ denote an experimental realization of the ideal one $\E$. By gate fidelities $\bar{F}$, we mean either the average gate fidelity \cite{ref:horodecki} \cite{ref:nielsen}, or the minimum fidelity \cite{ref:kribs},
\bea
\bar{F}_{\a} (\E,\E_{\x}) & = & \int d\psi F (\E[\psi], \E_{\x}[\psi]) \label{eq:avf},\\
\bar{F}_{\w} (\E,\E_{\x}) & = & \min_{\psi} F (\E[\psi], \E_{\x}[\psi]), \label{eq:minf}
\eea
where $F$ denotes Uhlmann's fidelity between quantum states \cite{ref:fidel}. Gate fidelities in the above estimate the similarity between resulting states by an experimental realization $\E_{\x}$ and by ideal one $\E$, respectively. We argue that these are appropriate to quantifying experimental realizations $\E_{\x}$ with respect to the ideal one $\E$. For gate fidelities $\bar{F} (\E,\V)$ between $\E$ and $\V$, the meaning is found in the fundamental aspect of quantum operations, i.e. the fundamental limitations in quantum theory to perform an operation $\V$. As an example, perfect cloning $\V$ is approximated by the optimal one-to-two universal quantum cloning $\E$ with fidelity $\bar{F} (\E,\V) = 2/3$ \cite{ref:no-go}. We emphasize that unless $\V$ can be perfectly done in quantum theory, by gate fidelities $\bar{F} (\E_{\x},\V)$ neither is shown how successful an experimental realization is, nor the optimality of quantum operations to perform $\V$.

To relate gate fidelities to quantum broadcasting, let us briefly introduce quantum broadcasting and shareability of bipartite quantum states. First, the notion of quantum state shareability reflects the fact that entanglement is a quantum resource that cannot be shared with arbitrarily many parties, also known as monogamy of quantum correlations \cite{ref:qshare} \cite{ref:doherty}. To be precise, a bipartite quantum state $\rho_{AB_{1}}$ is called $k$-shareable if an extended $k+1$ partite state $\rho_{AB_{1} \cdots B_{k}}$ can be found such that bipartite states $\rho_{AB_{i}}$ for all $i$ are identical, i.e. $\rho_{AB_{i}} = \rho_{AB_{j}}$ for $i \neq j$ where $\rho_{AB_{i}} = \tr_{\bar{A}\bar{ B_{i}}} \rho_{AB_{1} \cdots B_{k}}$. This means that when two parties $A$ and $B_{1}$ share a $k$-shareable quantum state, there can exist additional $k-1$ parties such that each of them shares the identical copy with $A$. This can also be used as an insecurity criteria in quantum key distribution \cite{ref:lutken}. A remarkable result along the line is that infinitely-shareable quantum states are only separable states \cite{ref:qshare}.


Next, the notion of quantum broadcasting generalizes quantum cloning \cite{ref:bdcasting}, and has an operational significance in its interpretation: information leaked out during the evolution can be quantified by the number of parties which learn about the resulting state of a given system. By saying that a quantum operation is $k$-broadcasting, we refer to a positive and completely positive linear transformations from a quantum state to a $k$-partite state such that quantum states of individual systems are identical. Note that a trace-preserving quantum operation is called a quantum channel. Throughout the paper, we also write $S(\H)$ to denote a set of quantum states over Hilbert space $\H$. Then, a quantum channel $\E^{(k)}:S(\H) \rightarrow S(\H^{\otimes k})$ is $k$-broadcasting if for any state $\rho \in S(\H)$, the resulting $k$-partite state, $\rho_{1,\cdots,k} = \E^{(k)}[\rho]$, fulfills that $\rho_i = \rho_j$ for $i\neq j$ where $\rho_{i} = \tr_{\bar{i}} \rho_{1, \cdots, k}$ \cite{ref:note1}. We then write $\E_{i}^{(k)}$ to denote a local mapping of an $i$th party as $\E_{i}^{(k)}: S(\H) \rightarrow S(\H)$, i.e. $\E_{i}^{(k)} =\tr_{\bar{i}} \E^{(k)}$.

A quantum operation for a given system corresponds to a local mapping. Once it is implemented, the realization can be completely identified by the so-called quantum process tomography (QPT) \cite{ref:qpt}. More precisely, a single matrix called the Choi-Jamiolkowski (C-J) state, denoted by $\chi_{\E}\in S(\H\otimes \H)$, is obtained: $\chi_{\E} = (\openone \otimes \E)[|\phi^{+} \rangle \langle \phi^{+}|]$ where $|\phi^{+}\rangle =\sum_{i}|ii\rangle /\sqrt{d}$, and gives the complete characterization of performed operations \cite{ref:C-J}. Conversely, for a channel $\E: S(\H) \rightarrow S(\H)$ identified by QPT, one can ask if there could be a $k$-broadcasting extension $\E^{(k)}$. If an operation $\E$ allows such extension, this means that there can be $k-1$ ancillary systems which can also learn about the resulting state of the system. Quantum operations for which no such extension is possible are, what we call, \emph{private operations} in the sense that the operation does not distribute system's resulting states to any other parties.



Having collected results in the above, we now show the correspondence between quantum broadcasting and quantum state shareability. \\

\emph{A quantum operation $\E :S(\H) \rightarrow S(\H)$ is a local mapping of a $k$-broadcasting operation $\E^{(k)} : S(\H) \rightarrow S(\H^{\otimes k})$  if and only if its C-J state $\chi_{\E}$ is $k$-shareable.} \\


Proof. First, suppose that $\chi^{AB_{1}}$ for an operation $\E_{B_{1}}$ is $k-$shareable, meaning that there exists an extension $\chi^{AB_{1}\cdots B_{k}}$ such that $\chi^{AB_{i}} = \chi^{AB_{j}}$ for all $i,j$. Let us now construct a channel using the extended state, $\E^{(k)} : S(\H) \rightarrow S(\H^{\otimes k})$,
\bea \E^{(k)} [\rho] = d_{A} \tr_{A} [\chi_{\E}^{AB_{1}\cdots B_{k}} (\rho^{T}\otimes \openone_{B_{1}} \otimes \cdots \otimes \openone_{B_{k}})]. \nonumber \eea Let $\rho_{1,\cdots,k}$ denote a resulting state, $\rho_{1,\cdots,k} = \E^{(k)} [\rho]$, for an input state $\rho$. Since $\chi^{AB_{1}\cdots B_{k}}$ is $k$-shareable, it holds that $\rho_{i} = \rho_{j}$ for all $i \neq j$ and therefore it is shown that $\E^{(k)}$ is $k$-broadcasting. Then, a local mapping, $\E_{i}$ of a party $B_{i}$ for any $i=1,\cdots, N$ can be found from $\tr_{\bar{i}} \E^{(k)}$, whose C-J state satisfies, $\chi^{AB_{i}} = \chi^{AB_{1}}$ for all $i$.

Conversely, let $\E_{i}$ denote the $i$th local mapping of a $k$-broadcasting channel $\E^{(k)}$, i.e. $\E_{i} = \tr_{\bar{i}} \E^{(k)}$. All of them are identical, $\E_{i} [\rho] = \E_{j} [\rho]$ for all $\rho$ and $i,j \leq k$. This leads to the equality, using the C-J isomorphism, $\E_{i} [\rho] = d_{A}\tr_{A}[\chi^{AB_{i}} (\rho^{T} \otimes \openone)]$, that \bea \chi^{AB_{i}} & = & \chi^{AB_{j}},~~ \forall i,j \leq k.\label{ref:cjs} \eea One can also find a multipartite state from the application of the $k$-broadcasting channel, $\chi_{\E}^{AB_{1}\cdots B_{k}} = [\openone\otimes \E^{(k)}](|\phi^{+}\rangle \langle \phi^{+}|)$, in which Eq. (\ref{ref:cjs}) holds. This completes the proof that $\chi_{\E}^{AB_{i}}$ is $k$-shareable. $\Box$\\

The result in the above establishes the equivalence between two operational quantities, quantum broadcasting and quantum state shareability, via the isomorphism between quantum states and operations in Ref. \cite{ref:C-J}. Based on this, results in quantum state shareability can be translated into quantum operations, and vice versa. As an example, let us consider infinitely- and $1$-shareable quantum states. Note that C-J states are actually obtained from QPT. That is, \emph{broadcasting properties of quantum operations are immediately found from the shareability of C-J states}.

First, as it was shown in Ref. \cite{ref:qshare}, only separable states are infinitely-shareable. As separable states correspond to entanglement-breaking channels via the C-J isomorphism \cite{ref:entbreak}, \emph{only entanglement-breaking channels can broadcast resulting states to $N$ parties for any $N$}. This also holds for asymptotic quantum cloning \cite{ref:acin}, the convergence of which can be found in Ref. \cite{ref:classical}. Next, it is clear that pure entangled states are $1$-shareable, i.e. for any bipartite entangled state $|\psi\rangle_{AB}$, there is no tripartite state $\rho_{ABC}$ such that $\tr_{C}\rho_{ABC} = |\psi\rangle_{AB}\langle \psi|$ and $\tr_{B}\rho_{ABC} = |\psi\rangle_{AC}\langle \psi|$. This means that for a given channel, if its C-J state is pure and entangled, the channel cannot be used for broadcasting of quantum states. Consequently, a unitary transformation can never be used for broadcasting since its C-J state is a maximally entangled state. It is thus shown that \emph{unitary transformations are private operations.}


\emph{Proposal to implement private operations.} Quantum operations to a given system are generally described by Kraus operators, $\{K_{i}:~K_{i}^{\dagger}K_{i}\geq 0,~~ \sum_{i}K_{i}^{\dagger}K_{i} = I\}$ such that, $\E[\rho] = \sum_{i} K_{i} \rho K_{i}^{\dagger}$ \cite{ref:Kraus}. If the evolution is not unitary, its C-J state forms a mixed state allowing an extension to extra parties, and consequently system's resulting states can be broadcasted to a number of parties.


The fact that unitary transformations are private operations can be used to implement approximate quantum operations as private operations. This can be done by applying the Stinespring representation of quantum operations \cite{ref:stine}: any quantum operation to a given system can be performed by unitary transformation over given and ancillary systems, i.e. there exist unitary $U_{SA}$ and ancillary state $|a\rangle_{A}$ such that $\E[\rho_{S}] = \tr_A U_{SA} \rho_{S}\otimes|a\rangle_{A} \langle a | U_{SA}^{\dagger}$. Since unitaries do not broadcast resulting quantum states, as long as ancillary systems are under control (e.g. stored in quantum memory), no other extra party is allowed to learn about system's resulting states. Therefore, any quantum operation implemented in the Stinespring form is private as long as ancillary systems are under control. This also signifies an operational difference between two alternative descriptions for quantum operations, Stinespring and Kraus forms.





\emph{No perfect cloning of unitaries.} The fact that unitary transformations cannot be applied for broadcasting of quantum states also leads to a corollary that a unitary transformation cannot be copied. To show this, suppose that perfect cloning of a unitary transformation $U$ is possible. This also means that there exists a C-J state, $\chi_{U \otimes U}$ for two copies of $U$, such that $\chi_{U} =\tr_{1}[\chi_{U \otimes U}] = \tr_{2}[\chi_{U \otimes U}]$ where $\tr_{i}$ denotes tracing out the $i$th system. As it was mentioned, however, $\chi_{U}$ is a pure entangled state and does not allow such an extension. This reproduces the result in Ref. \cite{ref:Guilio} that \emph{unknown unitary transformations cannot be perfectly cloned}. The same also holds true for quantum operations whose C-J states are pure and entangled.

\emph{Bounds to gate fidelities.} We now derive a general bound to gate fidelities translating the hierarchial structure among shareable quantum states, shown in Ref. \cite{ref:doherty}, into quantum operations. Let $S_k$ denote the set of $k$-shareable bipartite quantum states. For instance, $S_{\infty}$ corresponds to separable states and pure entangled states only belong to $S_1$. A natural relation of inclusion follows, $S_{\infty} \subset \cdots \subset S_{2} \subset S_{1}$, which has been used to derive a complete criteria for separability of quantum states \cite{ref:doherty} \cite{ref:migeul}. The trace distance, defined as $D(\rho,\sigma) = \|\rho - \sigma \|/2$ for states $\rho$ and $\sigma$, can be used to estimate distances between quantum states: the minimal distance of a state $\rho \in S_{k}$ to the set $S_{\infty}$ is bounded as follows,
\bea \min_{\sigma\in S_{\infty}} D ( \rho, \sigma)  \leq \epsilon (k), ~~ \mathrm{where} ~~\epsilon(1) > \cdots > \epsilon(\infty) =0.
\label{eq:bd}\eea
There have also been approaches to derive tighter bounds $\epsilon(k)$ \cite{ref:migeul}, \cite{ref:brandao}.


The hierarchy in quantum state shareability can be translated to quantum operations, via the inequality $F(\rho,\sigma)\geq  1 - D (\rho,\sigma)$ \cite{ref:inq}. Recall the relation between quantum operation $\E$ and its C-J state $\chi_{\E}$, $\E[\rho] = d\tr_A [\chi_{\E} (\rho^{T} \otimes I )]$ for a state $\rho$, where $d$ is the dimension of underlying Hilbert space. From Eq. (\ref{eq:bd}), it follows that for channel $\E$ and an entanglement-breaking one $\E_{EB}$,
\bea
D (\E[\rho], \E_{EB}[\rho]) \leq d  D(\chi_{\E}, \chi_{\E_{EB}}) \leq d \epsilon(k), ~ \forall \rho \in S(\H), \label{eq:inq}
\eea
where the C-J state $\chi_{\E_{EB}}$ belongs to $S_{\infty}$. Since $D$ is a distance measure, the triangle inequality holds:
\bea D (\E [\rho], \N[\rho]) \leq D(\E [\rho], \E_{EB}^{*}[\rho]) +D(\E_{EB}^{*}[\rho], \N[\rho]), \label{eq:tri}\eea
for some noise channel $\N$, where the channel $\E_{EB}^{*}$ satisfies: $D (\E [\rho], \E_{EB}^{*}[\rho]) = \min_{\E_{EB}} D (\E [\rho], \E_{EB}[\rho])$. Note that the noise channel is taken to estimate a worst realization of a given operation $\E$. Applying inequalities Eqs. (\ref{eq:inq}) and (\ref{eq:tri}) in the above, as well as the above-mentioned inequality of fidelity and distance, gate fidelities $\bar{F} (\E, \N)$ are bounded as follows,
\bea \bar{F} (\E, \N) \geq 1 - d \epsilon(k) - d\bar{D}(\E_{EB}^{*}, \N), \label{eq:M}\eea where $\bar{D}$ denotes gate distances, obtained by replacing fidelities in Eqs. (\ref{eq:avf}) and (\ref{eq:minf}) with trace distances.

Therefore, Eq. (\ref{eq:M}) provides a general lower bound to gate fidelities between a given operation and a noise one. Note that the distance $\bar{D}(\E_{EB}^{*}, \N)$ is generally small since a noise operation is very likely to be entanglement-breaking. Then, the bound depends on the shareability of the C-J state for a given operation, or equivalently, the broadcasting property of a given operation.

From this, we now show that a high value itself in gate fidelities does not suffice to assess experimental realization of quantum operations. Suppose that an approximate operation $\E$, say $k$-broadcasting, is implemented in experiment. It follows that $\bar{F}(\E, \E_{\x}) \geq \bar{F}(\E, \N)$ for a noise operation $\N$. From Eq. (\ref{eq:M}), no matter how successful or unsuccessful the realization is, gate fidelities are larger than $1- d\epsilon(k) -d \bar{D} (\E_{EB}^{*}, \N)$, which is close to $1 - d\epsilon(k)$ since $\bar{D} (\E_{EB}^{*}, \N) \ll 1$. That is, we have $\bar{F}(\E, \E_{\x}) \gtrapprox 1-d\epsilon(k) $: \emph{the more parties a given quantum operation can broadcast to, the higher gate fidelities of its experimental realization in general.}

\begin{figure}[t]
\includegraphics[width=3in]{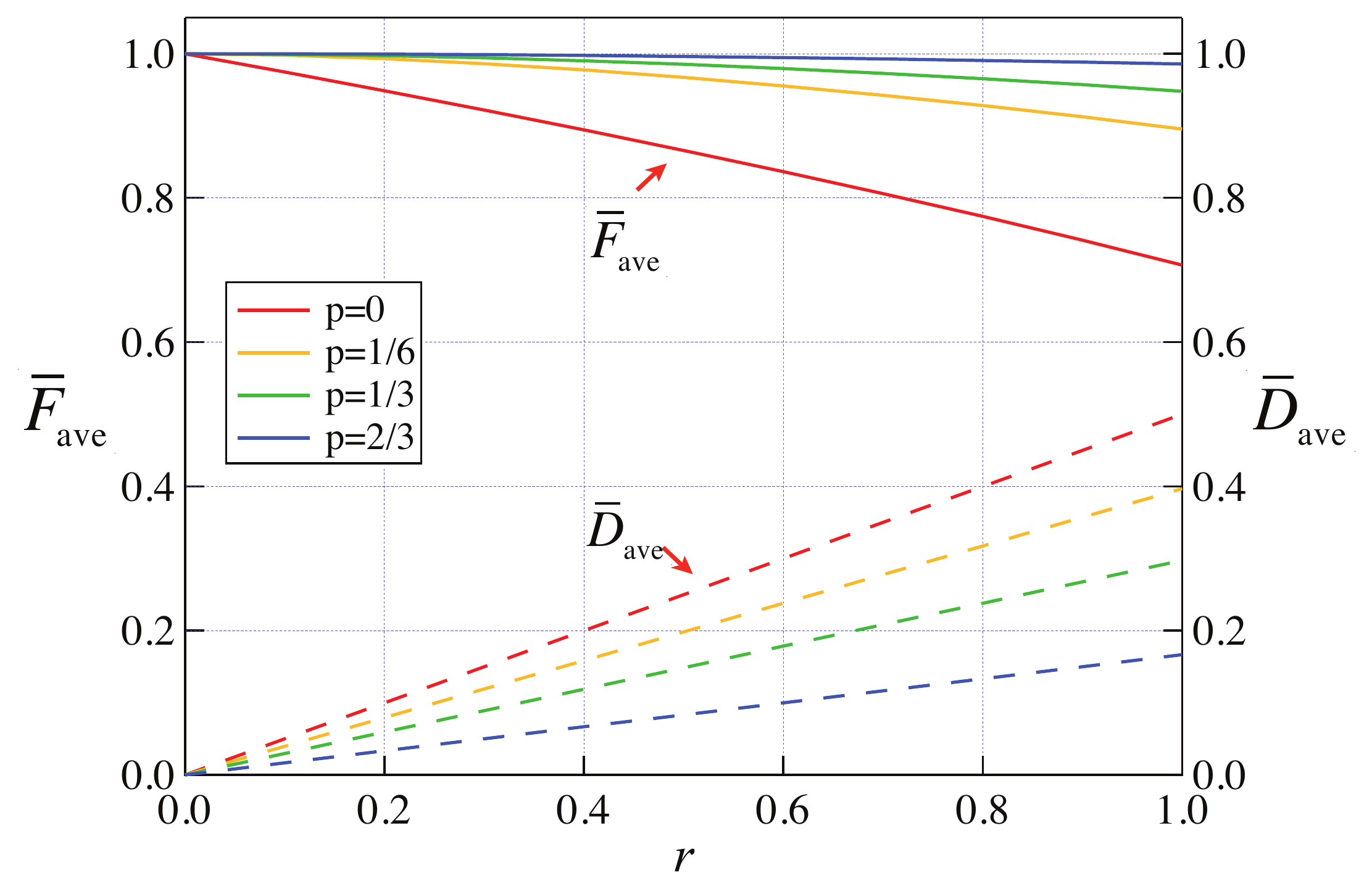}
\caption{The average gate fidelity and the average trace distance of the operation in Eq. (\ref{eq:pop}) are shown. }\label{fig:graph}
\end{figure}

Let us illustrate this with an example, $ \E_{p}[\rho] = (1-p)\rho + \frac{p}{2} (\sigma_{x}\rho\sigma_{x} + \sigma_{z}\rho\sigma_{z})$, which is entanglement-breaking for $p\geq 2/3$, and the case with $p=2/3$ corresponds to an optimal (approximate) transpose operation \cite{ref:T}. Suppose that the operation is realized for a given $p$ under the depolarizing noise,
\bea \E_{\x} [\rho]= (1-r)\E_{p}[\rho] + r(I/2). \label{eq:pop}\eea Larger values of $r$ imply more noise in realization. We compute $\bar{F} (\E_{p} , \E_{\x})$ in terms of the average fidelity, see Fig. \ref{fig:graph}. For ideal operation $\E_{p}$ with larger $p$ (i.e. broadcasting to more parties), the gate fidelity has higher values in general. For instance, when the ideal one is $\E_{p=0}$, the lowest value in the fidelity is about $0.70$. For $p=2/3$, no matter how noisy a performed operation is, the gate fidelity is immediately larger than about $0.98$.


\emph{Conclusions.} We have shown that for a given quantum operation, gate fidelities of its experimental realization are closely related to the broadcasting property of the operation. In fact, the broadcasting property gives a lower bound to gate fidelities of an experimental realization, \emph{independent to how successful or unsuccessful the realization is}. This is obtained by establishing the correspondence between two operational quantities, shareability of quantum states and broadcasting by quantum operations. The correspondence also enables us to translate results in quantum state shareability into broadcasting properties of quantum operations.

On the practical side of assessing experimentally realized quantum operations, we have learned that equal values in gate fidelities for different quantum operations cannot lead to the same conclusion. Moreover, high values in gate fidelities themselves cannot conclude successful realization of quantum operations. Nevertheless, as we have argued in the beginning, gate fidelities are appropriate measures in that they quantify how faithful experimental realization is. The results, therefore, suggest that, to assess an experimental realization of a quantum operation, a lowest bound to gate fidelities should also be presented taking into account the worst case of realization such as noisy operations e.g. the depolarization. In fact, gate fidelities provides information on how good an experimental realization is compared to the worst case of the realization, and how noisier it is than the ideal one in terms of the broadcasting property.


This work was supported by  National Research Foundation of Korea (2009-0070668, 2009-0084473, and  KRF-2008-313-C00185). J.B. thanks the hospitality of the Mittag-Leffler institute, where a part of this work was done.



\end{document}